\documentstyle[aps,preprint,epsfig]{revtex}
\begin{document}
\title{An Analysis of the Asymptotic Limit of Gluon Shadowing.}

\author{S. Liuti $^{(a)}$
\thanks{On leave from: INFN-Sezione Roma Tre, Dipartimento
di Fisica E. Amaldi, Via Vasca Navale, 84. 00146 Roma, Italy.} 
and  F. Cano $^{(b)}$  }

\address{$^{(a)}$ Institute of Nuclear and Particle Physics, University
of Virginia, \\ McCormick Road, Charlottesville, Virginia 22901, USA.
\\ 
$^{(b)}$ Dipartmento di Fisica, Universit\'a di Trento, \\
Via Sommarive 14, 38050 Povo, Trento, Italy.}

\maketitle       

\begin{abstract}
We examine the gluon distributions in nuclei   
in the asymptotic region defined by   
$Q^2 \rightarrow \infty$, $x \rightarrow 0$. An analysis 
using the Double Asymptotic Scaling 
variables of Ball and Forte  
is proposed. New scaling relations are predicted which can
help disentangling the different mechanisms of low $x$ perturbative 
QCD evolution in nuclei. 
  
\vspace{2cm}
\begin{center}
{\em Submitted to Physics Letters B}
\end{center}

\end{abstract}


\narrowtext

{\bf 1.}
Nuclear shadowing or the depletion at low Bjorken $x$
of the nuclear Deep Inelastic (DI) structure 
function, $F_2^A$, with respect to the nucleon one, $F_2^N$, has been 
observed in a number of experiments (\cite{Arneodo} and references therein). 
Assuming the universality of parton distributions in nuclei, 
one expects nuclear shadowing to be present in other 
high energy processes as well, 
such as Drell-Yan pair, $J/\psi$ and $\Upsilon$ 
production in lepton-nucleus, hadron-nucleus and nucleus-nucleus 
collisions (\cite{Leitch} and references therein). 
In particular nuclear shadowing might be concurring with the other 
mechanisms, among which is the quark-gluon plasma formation, 
that result in a depletion of the 
observed cross sections for these processes. 
%\cite{Gyu,Ramona,Mar99}
Moreover, as the very low $x$ regime has become accessible 
at HERA, new phenomenological studies of high density QCD 
at the saturation scale predicted in \cite{Mue99} are now 
possible \cite{Lev99}.   
A quantitative understanding of both the $x$ and $Q^2$ dependences of the
nuclear parton distributions at low $x$ therefore constitutes 
a practical and necessary step both for interpreting  
the outcome of future experiments at RHIC and at the LHC
and for investigating the onset of parton saturation. 

Recent calculations rely on   
non-perturbative models for 
the nuclear parton distributions at a given (low) scale, 
$Q_o^2$, combined with DGLAP \cite{DGLAP} perturbative evolution.
They are all therefore affected by the uncertainty in the initial parton
distributions and, in particular, in the gluon distribution which governs
evolution at low $x$, and which is poorly known experimentally. 
A strong effect is seen by changing the value of $Q_o^2$ itself which
can plausibly vary within the range, $Q_o^2 = 0.8 - few \, {\rm GeV^2}$,
leading to sensibly different values
for the shadowing of both the 
structure function and the gluon distribution at large $Q^2$.  

%% S
Now, as $Q^2$ increases the 
(low $x$) gluon distribution in a proton should tend 
to a universal asymptotic value, corresponding to the 
Double Leading Logarithmic Approximation (DLLA) result of   
Ref.\cite{derujulaetal}. Numerically, this value is attained 
at $Q^2 \lesssim 10^3 \, {\rm GeV}^2$ 
provided the initial distributions grow much slower than $\approx x^{-1/2}$. 
Because of the coupling between the singlet quark and gluon distributions' 
evolution, a similar behavior is predicted for the structure function, 
$F_2$ \cite{MRS88}.  
%A similar behavior is reflected into the structure function, $F_2$ \cite{MRS88}.  
It is therefore natural to address the question of whether 
the differences in the initial non perturbative  
nuclear shadowing will decrease with growing $Q^2$ and give rise to 
a universal asymptotic curve which is within the range of 
planned experiments. Our first result is a negative one:
we will show that because of the form that  
DGLAP evolution takes in a nucleus
the influence of initial conditions is carried on to the largest 
attainable $Q^2$ values.    

%
% END S   
%%%%
%%%% END SIMONETTA
%%%%    
%% Note to myself: HLS get independence from the inital conditions 
% in their model at large Q^2. This works because of the same mechanism 
% as here, i.e. the gluons go to their DLLA universal value. 
% The value that the ratio R_G reaches though is model dependent and in 
% principle Q^2 dependent becasue it is given by the ratio of derivatives
% in a nucleus and in a nucleon. They try to pass it as a fixed point thing
% but it is just the rule of derivation of ratios, it 
% can be very misleading....find counter example for longer paper. 
 
%%% S
We then examine carefully the asymptotic 
behavior of 
the shadowing ratios $R_G=G_A/G_N$ and
$R_F=F_2^A/F_2^N$, in the DLLA (notations are: 
$G_{N(A)}$ and $F_2^{N(A)}$ 
for the gluon distributions
and the structure function in a nucleon, $(N)$, and 
in an isoscalar nucleus, $(A)$, respectively). 
Our goal is to ascertain whether  
it is possible to distinguish among the different approximations 
the perturbation series takes in a nucleus and at very low $x$.  

The proton structure function data analyzed recently  
at HERA \cite{H1,ZEUS} have been shown \cite{H1}  
to lie in the asymptotic region
($Q^2 \gg Q_o^2=1 \, {\rm GeV^2}$ and $x \ll x_o \, = 0.1$, 
and to evolve according to DLLA \cite{BF_1}. 
%%%% END S
The key test is to prove that the data obey Double Asymtptotic Scaling (DAS) 
in the variables 
$\rho = \gamma ((Y-Y_o)/\xi)^{1/2}$ and 
$\sigma = \gamma^{-1}((Y-Y_o)\xi)^{1/2}$, $\gamma=6/(33-2N_f)^{1/2}$, 
$Y=\ln 1/x$, 
$\xi=\gamma^2\ln (\ln Q^2/\Lambda_{QCD}^2/\ln Q_o^2/\Lambda_{QCD}^2$).
Violations from DAS (other than due to the fact that the data lie in a 
pre-asymptotic region  \cite{BF_1}) 
would signal either the onset of contributions beyond standard pQCD evolution
\cite{DGLAP}, including 
the beginning of parton saturation \cite{Mue99,Lev99}. 

%
% S
The approach to asymptotia in a nucleus can
be first analyzed 
by assuming that evolution proceeds through 
DLLA equations as well.  
We have found  that in this case 
the ratios $R_F$ and $R_G$ display exact scaling in $\sigma$, thus becoming 
a function of $\rho$ only. As in the proton case, 
this is  
a model independent result in that we obtain a scaling form, independent from 
the initial conditions. 
We then use this result as a basis for addressing the next question {\it i.e.} 
the detection of violations of DAS scaling in nuclei, which could possibly 
originate  
%
% END S
at different
values of $x \equiv x_o^A$ and $Q^2 \equiv Q_{o,A}^2$, than in the proton.   
In particular Unitarity Shadowing  
Corrections (USC) \cite{GLR,MueQiu,Levinetal} 
are expected to affect evolution at  
$x_o^A > x_o$ because of the increase of the 
%%% S
tranverse 
%%% END S
gluon density in a nucleus
due to the overlapping of nucleons in the longitudinal direction. 
%This idea is supported by earlier numerical 
%calculations \cite{CK,EQW} which 
%indicate that in nuclei the saturation region 
%might be at reach of currently proposed experiments. 
%%% CHECK DONNACHIE
On a more speculative basis one might also expect the transition to the 
$\ln(1/x)$ resummation to appear in 
a different regime 
%
% S
or, in the most ``exotic'' scenario, that medium 
modifications of the anomalous dimensions could be observed.
%%%% END S
In our approach such questions can be addressed systematically as 
they introduce
specific {\em scaling violations} from the DLLA result, 
appearing as different $\sigma$ dependences in the ratios $R_G$ and $R_F$.

%%%% S
Our main observation is therefore that although it is technically 
predictable that for a proton target 
DGLAP evolution and DLLA should break down
at very low values of $x$ and sufficiently large $Q^2$
and give way to $\ln (1/x)$ summation and to USC,    
it is still a major task to be able to pinpoint where and if the 
transition from the different regimes is going to take place
in the kinematical regimes currently under exploration.
%%%%END S
Our goal is to obtain some new insight  
by using nuclear targets where the asymptotic regime
can in principle be reached at larger $x$. 
As a by-product we obtain quantitative predictions in the 
asymptotic kinematic regime which should be attainable at RHIC and at 
the LHC. 
         
{\bf 2.}
We first summarize results for ordinary DGLAP evolution applied to the nuclear
ratios at low $x$, 
assuming that the proton and the nuclear distributions evolve
similarly. 
As it is well known evolution is driven by the gluon distribution 
which dominates over the
sea quarks one and  
one can predict the behavior of the shadowing 
ratios, $R_G$ and 
$R_F$ with $Q^2$:
\begin{eqnarray}
\frac{\partial R_G }{\partial \ln Q^2} & \simeq & 
 \int_x^1     
P_{GG}\left(\frac{x}{y},\alpha_S(Q^2)\right) \frac{G_N(y,Q^2)}{G_N(x,Q^2)}
\left[ R_G(y,Q^2)- R_G(x,Q^2)\right] \frac{dy}{y} 
 \nonumber \\
& \equiv &  \frac{\partial G_N/\partial Q^2}{G_N}
\left( \frac{\partial G_A(x,Q^2)/\partial \ln Q^2}
                 {\partial G_N(x,Q^2)/\partial \ln Q^2} - R_G(x,Q^2) \right), 
\label{evol}
\end{eqnarray}
\begin{equation}
\frac{\partial R_F}{\partial \ln Q^2}  \simeq 
\int_x^1  
P_{qG}\left(\frac{x}{y},\alpha_S(Q^2)\right) 
\frac{G_N(y,Q^2)}{\Sigma_N(x,Q^2)} 
\left[ R_G(y,Q^2)- R_F(x,Q^2)\right] \frac{dy}{y},   
\end{equation} 
\noindent 
where we have disregarded the sea quarks distribution on the 
{\it r.h.s.} of the coupled DGLAP evolution equations; 
$P_{qG}$ and $P_{GG}$ are the splitting functions 
evaluated at NLO; and we used the approximation 
$F_2^{N(A)} \approx 5/18 \Sigma_{N(A)}$, with $\Sigma= 
\sum_i q_i(x,Q^2)+ \bar{q}_i(x,Q^2)$.
For ease of presentation we will use the following notation: 
$G_{N(A)}^\prime = \partial G_{N(A)}(x,Q^2)/\partial \ln Q^2$.
Eqs.(1) and (2) show that the $Q^2$ dependence of the ratios $R_G$ and $R_F$ is 
determined by a subtle balance involving both the parton distributions 
and the ratios themselves \cite{Qiu}. 
%%%%%
%%%%% S
Based on Eqs.(1) and (2), and 
defining $R_G(x,Q_o^2) \equiv R_G^o$ and $R_F(x,Q_o^2) \equiv R_F^o$
for the initial distributions, one can make the following predictions for 
the $Q^2$ dependence of $R_G$ and $R_F$: 
{\it i)} $R_G$ grows with $Q^2$.
In fact, if as predicted by non-perturbative shadowing 
models
$R_G$ is a growing function of $x$, then it also grows with $Q^2$, the {\it r.h.s.}
of Eq.(1) being positive ($y \geq x$);  
{\it ii)}
if $R_F^o < R_G^o$, 
then $R_F$ grows with $Q^2$; if  
$R_F^o > R_G^o$, then $R_F$ initially decreases with $Q^2$ until 
it reaches the value of $R_G^o$ and it subsequently starts increasing 
along with $R_G$.  

Note that from the behavior of $R_F$ and $R_G$ 
obtained from the straighforward application of DGLAP equations,  
one cannot predict the approach of {\it e.g.} $R_G$ to a universal limiting curve
at large $Q^2$. As a matter of fact,
although the form of Eq.(1) might seem 
suggestive of a fixed point behavior \cite{HLS}, 
this is {\it a priori} not the case, since 
the quantity $G_A\prime/G_N\prime$ depends on $Q^2$. 
%
% END S
The rate of change with $Q^2$ is 
instead governed both by: (a) by the ratio, $G_N^\prime/G_N$;
(b) by the difference 
$\Delta_G=G_A^\prime/G_N^\prime - R_G(x,Q^2)$ at $Q^2=Q_o^2$ .
Current parametrizations \cite{MRS98,GRV98} feature an ultra-soft behavior 
of the gluon distribution at $Q_o^2 \leq 1 \, {\rm GeV^2}$, {\it i.e.} 
$G_N \rightarrow 0$ as $x \rightarrow 0$, 
thus causing (because of (a)) a rapid evolution which strongly reduces the shadowing 
in both $R_G$ and $R_F$, 
between $Q_o^2$ and $Q^2 \approx 1-2 \, {\rm GeV^2}$.  
If on the contrary one assumes $Q_o^2$ ranging from $2$ to $5 \, {\rm GeV^2}$, 
where harder low $x$ initial gluons are expected, 
then the evolution is slower ($G_N$ is larger) and the nuclear ratio 
is basically unchanged at 
$Q^2 \approx 10-100 \, {\rm GeV^2}$.
A further model dependence follows from the usage of different non-perturbative
shadowing mechanisms. We examine two in particular: the Aligned Jet Model (AJM)
(see \cite{FS} and references therein), and Initial State Recombination (ISR) \cite{CQR}.
Both models explain qualitatively the initial onset of shadowing. Accurate 
quantitative calculations have been performed using the AJM in \cite{Esk,FSL}.   

%% EXPLAIN %%%

%FLORENCIO's ADDITION
Results are
summarized in Fig.\ref{DGLAP}, where we show the $Q^2$ dependence of the ratios
$R_F(x,Q^2)$ and $R_G(x,Q^2)$ in $^{40}Ca$ at fixed $x=10^{-4}$ (Fig. 1(a)) and $x=10^{-2}$
(Fig. 1(b)), for both the AJM (full lines) and the ISR model (short dashes).  
In order to show the dependence on the initial scale
$Q_0^2$, results are presented for both models by taking, $Q_0^2=0.8$ GeV$^2$,
and $Q_0^2=5$ GeV$^2$ (the latter can be easily distinguished in the graph by 
observing the shift in the starting point of the curves). 
Moreover, as the main purpose of the figure is to illustrate the 
main features of both the nuclear models and the initial parton distributions 
that will lead to our description on the asymptotic behavior, 
we have not sought for $R_F$ the best agreement with the data. Details 
on this part are going to be given elsewhere \cite{CanLiu}.   
A remarkable feature is the sensitivity of the ISR model 
to this initial scale $Q_0^2$. It is only for 
$Q_0^2 \lesssim 1$ GeV$^2$ that a 
sizeable gluon shadowing is obtained due to the fact 
that the amount of initial shadowing is proportional to the square of the
initial gluon distributions and to $\alpha_s(Q_0^2)/Q_0^2$ \cite{CQR}. 
Results obtaind with the AJM vary less dramatically
with the initial scale, $Q_0^2$, which in this case enters just the $(q \bar{q})-nucleon$ 
(or $(gg)-nucleon$) cross sections \cite{FS}. 
Nonetheless, the initial difference is carried on to $Q^2$ as large as 10$^3$  GeV$^2$.
Moreover, in both cases we can observe the $Q^2$ behaviour outlined before: a 
rapid suppression of shadowing in the range of $Q^2$ up to 2 GeV$^2$ and a subsequent softer evolution. 
The comparison between the $Q^2$ behaviour for different fixed $x$ values (Fig. 1(a) and Fig. 1(b)) 
shows that evolution is slower for smaller $x$. This can be technically understood by noting that, 
in the DLLA limit the logarithmic derivative of $R_G$  in Eq. (\ref{evol})
vanishes. This is why initial discrepancies between models are more 
likely to persist at smaller $x$.

%%%%%
%In all cases examined here, presented in Fig.\ref{DGLAP} 
%evolution brings an attenuation of the 
%$Q^2$ dependence, quantitatively seen as a decrease
%in the slope of $R_G$. As we will explain later, $R_G$ is actually 
%zero in the asymptotic regime.      
In summary, the asymptotic behavior of $R_G$ depends on the initial conditions
up to the largest values of $Q^2$ attainable at low $x$.  
This feature is in common with the proton gluon distributions 
themselves as shown {\it e.g.} in \cite{MRS88}. 
%In Fig.1 we show also the results for $R_F$: note, in particular 
%the two different behaviors, {\it i.e.}  explain, 
%described in {\it ii)}.  
As it is well known, neither the data nor theoretical arguments 
can help us defining the optimal values of $R_G(x,Q_o^2)$ and $R_F(x,Q_o^2)$.

The situation that we have described calls for some redefinition  
of the approach to asymptotia in deep inelastic scattering from nuclei.
In the next Section we illustrate how different behaviors of the 
data could be revealed by extending the double asymptotic scaling
analysis of Ref.\cite{BF_1} to nuclei.      
  
% For me: Also, why is it that one has np gluon shadowing when 
% the initial gluons are so few? Is it a relative effect that does not depend 
% on the absolute value of the gluon density?   

%A quantititative understanding of the effects described in this Section 
%is especially important for 
%quarkonia production where $Q^2 \equiv M^2_{J/\psi} 
%\approx 10 \, {\rm GeV^2}$ and $Q^2 \equiv M^2_{\Upsilon} \approx 100 \, 
%{\rm GeV^2}$ can be both considered asymptotic if the initial scale is 
%$Q_o^2 \lesssim 1 \, {\rm GeV^2}$. 

%
% SECTION 3
%
{\bf 3.}
We now examine the nuclear DI structure function and gluon distribution 
in the asymptotic regime defined by $x \rightarrow 0$, $Q^2 \rightarrow \infty$.
Our goal is to explore scaling relations in nuclei in order to be able
to compare theory with data in a model independent way.
The derivation of the equations of the DLLA in a nucleus parallels 
the one for the proton, namely one first writes   
the DGLAP evolution equation for the gluon distribution in the limit 
$n \rightarrow 1$, $n$ being the variable in moments space 
(we have omitted the subscripts $N(A)$ unless necessary):
\begin{equation}
\frac{\partial g(n,Q^2)}{\partial \ln Q^2}   =  \frac{\alpha_s(Q^2)}{2 \pi} 
\gamma^0_{GG}(n) g(n,Q^2), 
\label{evoln}
\end{equation}
$\gamma^0_{gg}(n) \approx 2 C_A/(n-1) +\kappa$ being the anomalous dimension in 
the limit $n \rightarrow 1$ ($\kappa = -11/6-n_f/3C_A$ is the next-order or subleading 
contribution in this limit). 
Solutions in $(x,Q^2)$ are found by evaluating the anti-Mellin transform, 
\begin{equation}
G(x,Q^2) =  \frac{1}{2 \pi i} \int_C dn \, g(n,Q^2) 
\exp\left[ Y (n-1) \right],  
\label{DLLA0}
\end{equation}
with the saddle point method \cite{derujulaetal}. In Eq.(\ref{DLLA0}), 
$g(n,Q^2)= g(n,Q_0^2) \exp\left[\xi/(n-1) + \xi\kappa/2 \right]$; 
$g(n,Q_0^2) = \int_0^1 dx \, x^{(n-1)} g(x,Q_0^2)$, $g(x,Q_o^2)$ being the
initial gluon distribution, and $G(x,Q^2)=xg(x,Q^2)$;    
$Y=\ln(1/x)$, and 
$\xi= \gamma^2 \ln (\ln(Q^2/\Lambda^2)/(\ln Q_o^2/\Lambda^2))$, 
$\gamma^2 = C_A/(\pi b)$, $b=(33-2N_f)/12\pi$.  
We rewrite the integrand in Eq.(\ref{DLLA0}) as:
\begin{equation} 
\widetilde{g}(n,Q^2)=g(n,Q_o^2)\exp\xi\kappa/2 \exp[Y f_1(n)+\xi f_2(n)],
\label{integrand}
\end{equation}
where $f_1(n)=n-1$, $f_2(n)=(n-1)^{-1}$ and $Y$ and $\xi$ are both similarly large.
If one takes  ``soft'' initial conditions such as, 
$G(x,Q_0^2) \approx A_N x^{-\lambda}$, $\lambda \lesssim 0$, 
then $g(n,Q_0^2) \approx A_N/(n-(\lambda+1))$ has a pole to the left of 
the saddle point which is found 
by setting $\partial [Y f_1(n)+\xi f_2(n)]/\partial n =0$. 
The integral in Eq.(\ref{DLLA0}) is then approximated by:
\begin{eqnarray}
G(x,Q^2) &  =  & \sqrt{2 \pi}  
\left( \frac{\widetilde{g}(n_o,Q_o^2)}
{\widetilde{g}''(n_o,Q_o^2)} \right)^{1/2}
\widetilde{g}(n_o,Q^2) =   \nonumber \\
& = &  \sqrt{2 \pi} g(n_o,Q_o^2)\exp(\xi\kappa/2) 
\frac{\exp[Yf_1(n_o)+\xi f_2(n_o)]}{\left| Yf_1''(n_o)+\xi f_2''(n_o)\right|^{1/2}}
\label{DLLA1}
\end{eqnarray}
where 
$n_0= 1 + \sqrt{\xi/ \Delta Y}$, $\Delta Y = Y- Y_o$, $Y_o=\ln(1/x_o)$, $x_o \approx 0.1$, 
is the saddle point and 
$f_{1(2)}''(n_o) = 
\left.  \partial^2 f_{1(2)}(n) / \partial n^2 \right|_{n=n_o}$.

By introducing the variables, $\rho = \gamma ((Y-Y_o)/\xi)^{1/2}$ and 
$\sigma = \gamma^{-1}((Y-Y_o)\xi)^{1/2}$, \cite{BF_1} one has:
\begin{equation}
n_o \equiv n_o(\rho)= 1 + \gamma/\rho,
\label{saddle}
\end{equation}
and,
\begin{equation}
G \equiv G^{DAS}(\rho,\sigma) = \sqrt{\pi} f_G(\rho/\gamma) \left( \frac{\gamma}{\rho} \right) 
\frac{\exp \left[ 2\gamma\sigma - \gamma^2 \kappa/2 \right]  }{\sqrt{\sigma\gamma}},
\label{GDAS}
\end{equation}
where $f_G(\rho/\gamma)$ is a smooth function describing the initial conditions.

DAS is the prediction that, in the hypothesis of soft initial conditions, 
and in the asymptotic limit defined by $\sigma \rightarrow \infty$ 
and $\rho \approx O(1)$,
$\ln(G^{DAS}/f_G(\rho,\sigma))$ becomes a  
linear function of $\sigma$, independent of the value of $\rho$, 
with slope fixed by the known constant, $\gamma$.
  
A similar behavior is found for $F_2^p$ \cite{BF_1} and can therefore be compared 
to the available data \cite{H1,ZEUS}. The structure function's asymptotic behavior 
is in fact obtained by solving the equation \footnote{We refer here to the singlet
part of the structure function}:
\begin{equation}
\frac{\partial F_2^p(x,Q^2)}{\partial \ln Q^2} = 
\frac{5}{18} \frac{\alpha_S}{\pi} G(x,Q^2),
\label{F2DLA} 
\end{equation}
yielding: 
\begin{equation}
F_2^p(\rho,\sigma) = f_\Sigma(\rho,\sigma) \exp(2\gamma\sigma),
\end{equation}
where $f_\Sigma$ was derived by using the LO expression for $\alpha_s$ 
and, as for the gluons, it depends on the initial conditions.
The scaling of $F_2^p$ can be seen from Fig.\ref{kin} (top-right).
From Fig.\ref{kin} one can also see that the data from NMC (triangles) \cite{NMC}, 
corresponding to larger $x$ with respect to the 1995 HERA ones (open squares) \cite{H1}, 
as well as some of the more recent HERA data with very low $x$ and 
$Q^2$ (open dots) \cite{ZEUS}, violate scaling (for a better reading 
compare with top-left). 
These scaling violations have been interpreted as due to the fact that the 
kinematics is not yet asymptotic,
as it can be easily seen from the fact that the data lie well below $\sigma \approx 1$.    

Deviations from DAS can be predicted also in the case of  hard initial 
conditions
{\it i.e.} when $g(n_0,Q_0^2) \approx A_N/(n-(\lambda+1))$ with 
$\lambda \gtrsim 0.2$. 
Intuitively this corresponds to taking the limit $Y \gg \xi$ in Eq.(\ref{DLLA0}), thus
defining the new saddle point: $n_o=(1+\lambda) +1/\Delta Y$. The corresponding
gluon distribution is then
\begin{eqnarray}
G(x,Q^2) & = & f_G^h \exp \left[ \lambda (Y-Y_o) + \xi/\lambda +\xi\kappa/2 \right] 
\nonumber \\
& = & f_G^h \exp \left[  \lambda \sigma \rho  + (\gamma^2/\lambda +\kappa/2)\xi \right]   
\label{HardP}
\end{eqnarray}
with $f_G^h(\rho,\sigma)= \sqrt{2\pi\rho\sigma} A_N$ (see also \cite{CK}).
This behavior supports the presence of USC appearing as a non-linear term
in the evolution equation \cite{GLR}, which has the effect of damping 
the steep rise of the gluon distribution at small $x$.
An alternative possible explanation of DAS violations in the recent HERA data
\cite{ZEUS} which accounts also for the peculiar stooping of the logarithmic slope 
of the proton structure function, $\partial F_2/ \partial \ln{Q^2}$, 
at low $x$ and $Q^2$,   
is that indeed 
USC \cite{Levinetal} need to be taken into account.

We now study these two different scenarios
for the asymptotic behavior in nuclei, 
where it is well known that some aspects of perturbative 
evolution such as USC, are
expected (simply based on geometrical arguments) to arise at larger values of $x$.  
In Fig.\ref{kin} we present the world low $x$ data on the nuclear ratios,
Eq.(\ref{evol}) as a function of 
the DAS variable, $\sigma$ (bottom-right), and we compare both data and their 
kinematics (bottom-left) with 
the proton ones (top). We use in both cases the following 
values of  $Q_o^2$, $\Lambda_{QCD}$, 
$\rho$ and $\sigma$:
$Q_o^2 = 1 \, {\rm GeV}^2$, 
$\Lambda_{QCD}=185$ MeV (LO).  
\footnote{In the NLO analysis performed in \cite{BF_1} it was found that: 
$Q_o^2 = 1.8 \, {\rm GeV}^2$, 
$\Lambda_{QCD}=200$ MeV (NLO), $\rho \approx 1$ and $\sigma \gtrsim 1.2$.}  
From the figure it appears that the 
existing nuclear data are scarse and do not presently support a 
DAS type analysis and, moreover, they seem to lie mainly in 
a pre-asymptotic region. 
Experiments at RHIC and LHC are however expected to be able to cover 
the asymptotic region. Although experimental extractions of
the logarithmic slopes in nuclei 
have been performed in \cite{Arneodo}, very little can be concluded 
from these data as well \cite{CanLiu}.  
 
We evaluate the ratios $R_G$ and $R_F$, Eq.(\ref{evol}) in DAS, {\it i.e.}   
within the hypotheses:  
{\it i)} the quark and gluon distributions are initially shadowed due 
to some non-perturbative mechanism;
{\it ii)} the pQCD evolution mechanism is not
affected by the nuclear medium; 
{\it iii)} the initial distributions are soft.
The DLLA predictions are:
\begin{equation}
R_G(x,Q^2) = \left[\frac{\widetilde{g}_A(n_o,Q^2)}{\widetilde{g}_N(n_o,Q^2)} 
\right]^{3/2}
\left.[ \frac{\widetilde{g}_N''(n_o,Q^2)}{\widetilde{g}_A''(n_o,Q^2)} 
\right]^{1/2} ,
\label{RG_DLA}
\end{equation} 
where $n_o$ is the saddle point defined by Eq.(\ref{saddle}) for
both the proton and the nucleus 
We rewrite Eq.(\ref{RG_DLA}) as a function of the DAS 
variables by using Eq.(\ref{GDAS}):
\begin{equation}
R_G^{DAS} = \frac{f_G^A(\rho/\gamma)}{f_G^N(\rho/\gamma)}
\label{RG_DAS} 
\end{equation}
{\it i.e.} the exponential terms appearing in 
$G^{DAS}(\rho,\sigma)$, Eq.(\ref{GDAS}), 
are the same in a nucleus and in a single nucleon respectively,   
thus canceling the $\sigma$ dependence in the ratio $R_G$: 
$R_G$ is predicted to scale exactly in $\sigma$ to a smooth function
of $\rho$ that is determined entirely by the (soft) initial conditions. 
%%%

The onset of a different evolution 
mechanism in the nucleus will appear as a $\sigma$-{\em scaling violation} 
modifying the exponential behavior of Eqs.(\ref{GDAS})-(\ref{RG_DLA}) with
respect to the single nucleon case. 
Since the low $x$ behavior of recent HERA data seem to 
show evidence for rather large screening corrections, and at the same
time they do not rule out the hard pomeron contribution, we consider the
effect of:      
{\bf (A)} USC combined with soft initial conditions; 
{\bf (B)} USC combined with hard initial conditions; 
{\bf (C)} Hard initial conditions in both nucleon and nucleus, no USC.
      
The effect of USC is taken into account through a ``damping factor'', $D_G(x,Q^2) \leq 1$,  
(Ref.\cite{Levinetal} and references therein), evaluated  
using Mueller-Glauber's eikonal approximation \cite{Mue90}.  
We extended the calculation to nuclei by assuming 
that two gluons inside a nucleus are correlated by a larger 
confinement radius, $R_A\approx r_0 A^{1/3}$, 
than in a nucleon, corresponding to a smeared impact parameter 
space two-gluon form factor 
(details of this calculation will be included in \cite{CanLiu}). 
As a result the effect USC is enhanced  
in a nucleus with respect to a nucleon target, 
due to the larger transverse gluon density at similar values
of $x$.      
  
The asymptotic gluon distribution function is written in terms of the damping
factor as:
\begin{equation} 
G^{SC}_{N(A)}(\rho,\sigma)  =  D_G^{N(A)} \times G^{DAS}(\rho,\sigma) . 
\label{GSC} 
\end{equation} 
%& \equiv & \exp \left[ -\widetilde{D}_G(\rho,\sigma) \right],
%where $\widetilde{D}_G = ln(1/D_G)$ (a similar expression holds for $F_2$). 
%
By using Eq.(\ref{GSC}) We can now calculate the ratio $R_G$, for the cases listed above. 
We obtain:
\begin{mathletters}
\begin{eqnarray}
R_G^{(A)} (\rho,\sigma) & = &  
R_G^{DAS}(\rho) \times \frac{ \exp \left[ 2\gamma(\sigma-\sigma_A) - 
\frac{\gamma^2\kappa}{2} \frac{\sigma}{\rho}  \right]}
{ \exp \left[ 2\gamma(\sigma-\sigma_N) - 
\frac{\gamma^2\kappa}{2} \frac{\sigma}{\rho}  \right]} \equiv  \nonumber 
\\
&  \equiv & R_G^{DAS}(\rho) \times \exp \left[- 
(\sigma_A -\sigma_N) \right ] .
\\
R_G^{(B)}(\rho,\sigma) & = &
\frac{f_G^{h,A}}{f_G^{h,N}} \times 
\frac{\exp \left[ \lambda_A \sigma\rho + \frac{\gamma^2}{\lambda_A}  
\frac{\sigma}{\rho} - \sigma_A \right] }
{\exp \left[ \lambda_N \sigma\rho + \frac{\gamma^2}{\lambda_N} 
\frac{\sigma}{\rho} - \sigma_N \right]} \equiv \nonumber 
\\ 
&  \equiv & C \exp \left[ (\lambda_A- \lambda_N) \sigma\rho
+ \gamma^2(\frac{1}{\lambda_A} - \frac{1}{\lambda_N} ) \frac{\sigma}{\rho} \right] 
 \times \exp \left[- (\sigma_A -\sigma_N) \right ] .
\\
R_G^{(C)} (\rho,\sigma) & = & C \exp \left[ (\lambda_A- \lambda_N) \sigma\rho
+ \gamma^2(\frac{1}{\lambda_A} - \frac{1}{\lambda_N} ) \frac{\sigma}{\rho} \right] .
\end{eqnarray}
\end{mathletters}
Here $R_G^{DAS}$ is the same as in Eq.(\ref{RG_DAS});
$\sigma_{N(A)}(\rho,\sigma) = 1/2 \gamma \ln (1/D_G)$; and the shadowing for
the initial hard distributions has been parametrized as $R_G^{(o)} =f_G^{h,A}/f_G^{h,N}
= C x^\alpha$, with $C \approx 1.3$ and $\alpha \approx 0.08-0.1$.
Moreover, we have chosen $Q_o^2 = 1 \, {\rm GeV^2}$ for both soft and hard
initial conditions and $\lambda_N=0.35$. 
Our scaling result, Eq.(\ref{RG_DAS}), and the scaling violating ones,
Eqs.(15a)-(15c), are shown in Fig.\ref{scaling}
for two different values of $\rho$: $\rho=1.8$ well inside the 
asymptotic region shown in Fig.{\ref{kin}, and $\rho=3.4$ corresponding to
very low $x$ and almost fixed $Q^2$ where we expect 
standard DGLAP to break down. 

A few  comments are in order. 
Starting from soft initial conditions one obtains
either the $\sigma$-scaling curves (full lines) or the $\sigma$-scaling violation, 
Eq.(15a), induced by USC (dashed lines). 
These corrections are driven by the damping factor given for each value 
of $\rho$ by the dashed curves below. The   
decreasing trend with growing $\sigma$ is larger at large 
$\rho$ because of the correspondingly decreasing values of $x$ at similar 
values of $Q^2$.
On the other side, hard initial conditions, Eq.(\ref{HardP}) and dot-dashed curves
in Fig.3, show sensible deviations from $\sigma$-scaling
at large $\rho$ where the $\ln (1/x)$ term dominates over the 
$\ln Q^2$ one. This effect is enhanced by USC (dotted curves). 
An interesting observation is the change in slope of the scaling violations
both when passing from soft to hard initial conditions at low $\rho$, 
and when passing from low $\rho$ to large $\rho$.  
These and similar other regularities could be studied systematically 
both for the gluon and the structure function ratios once a 
much larger set of data will be available.          

Most importantly, with the approach proposed here we eliminate the ambiguities 
in the determination of the value of gluons and quarks nuclear 
shadowing illustrated in Fig.1, in that we identify scaling relations 
that must be verified independently from the initial non-perturbative 
nuclear shadowing.   

In conclusion, by applying DAS to nuclei 
we have shown model independent predictions, {\it i.e.}  
scaling in the variable $\rho$  
for the ratios of the nuclear gluon distributions
to the free nucleon ones in the asymptotic region. 
Similar relations hold
for the nuclear structure functions. 
We have considered a few possible sources of scaling violations due to the
onset of USC and to the domination of a hard pomeron.  
As a result, by studying the $A$-dependence of  
shadowing we find some new constraints
on perturbative evolution at low $x$.   
Our calculations are relevant for the regime accessible
at future experiments at RHIC, LHC
and at the $eA$ project at DESY \cite{eA}.

%%%%%%%%%%%%%%%%%%%%%%%%%%%%%%%%%%%%%%%%%%%%%%%%%%%%%%%%%%%%%%%%%%%%%%%%%%%%%%%%%%%%%%%
%%%%%%%%%%%%%%%%%%%%%%%%%%%%%%%%%%%%%%%%%%%%%%%%%%%%%%%%%%%%%%%%%%%%%%%%%%%%%%%%%%%%%%%

%\begin{figure}[htb]
%\vbox{
%\hskip.6truecm\epsfig{figure=qcd98_fig1.eps,width=6.truecm}}
%\vskip-.6truecm
%\caption{The value of $\tau^2$ vs. $n$ 
%extracted according to different models
%for the $n$-dependent coefficient $a_n(1)$ as explained in the text.}
%\vskip-.8truecm
%\label{fig1}
%\end{figure}

\begin{figure}[htb]
\vbox{
\hskip0.6truecm\epsfig{figure=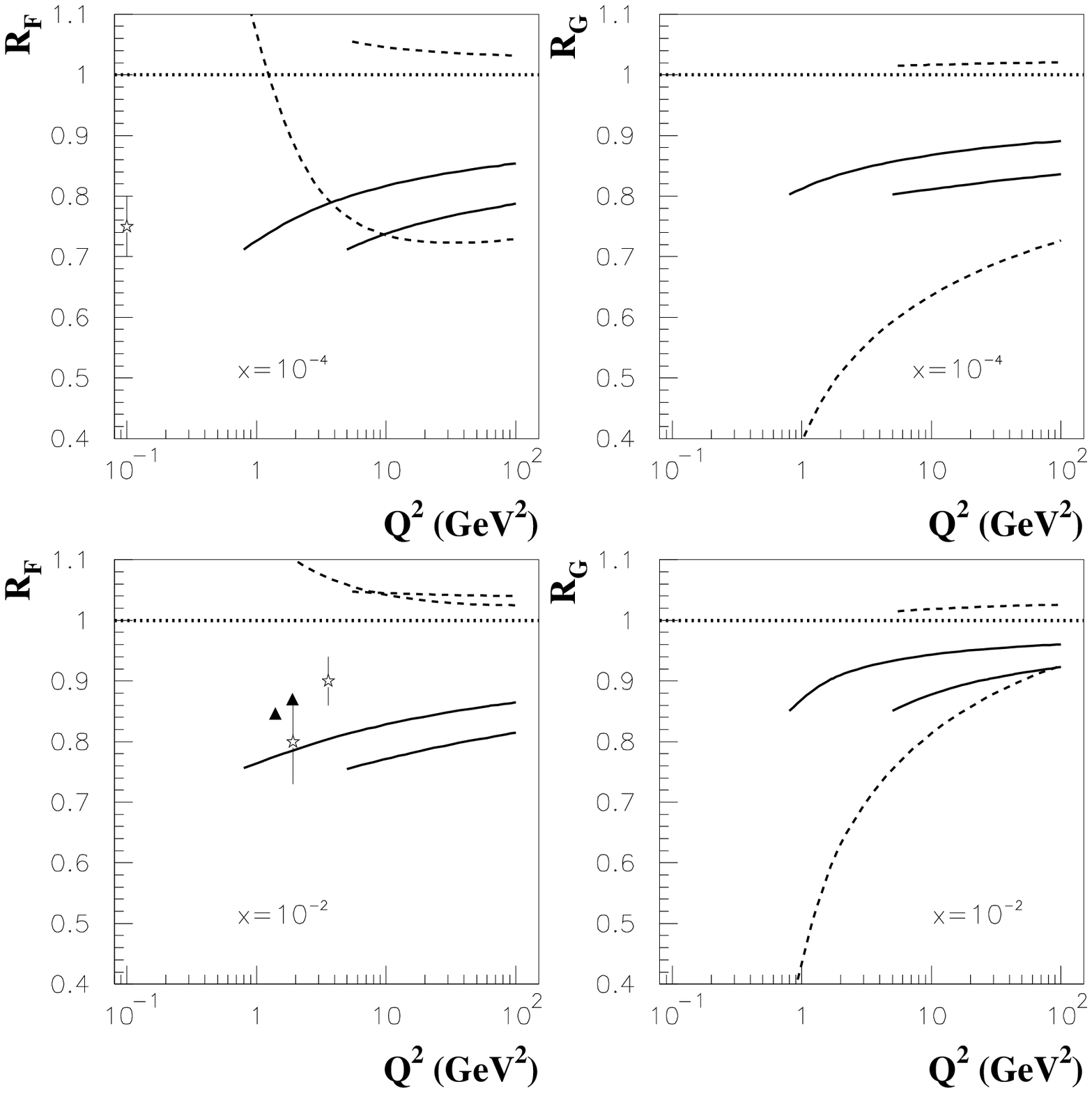,width=14.truecm}}
\vskip0.6truecm
\caption{DGLAP evolution in $^{40}Ca$. Experimental data
from NMC \protect\cite{NMC1} (triangles) and E665 \protect\cite{E665} 
(stars); theoretical curves are explained in the text.}
\label{DGLAP}
\end{figure}

\begin{figure}[htb]
\vbox{
\hskip.6truecm\epsfig{figure=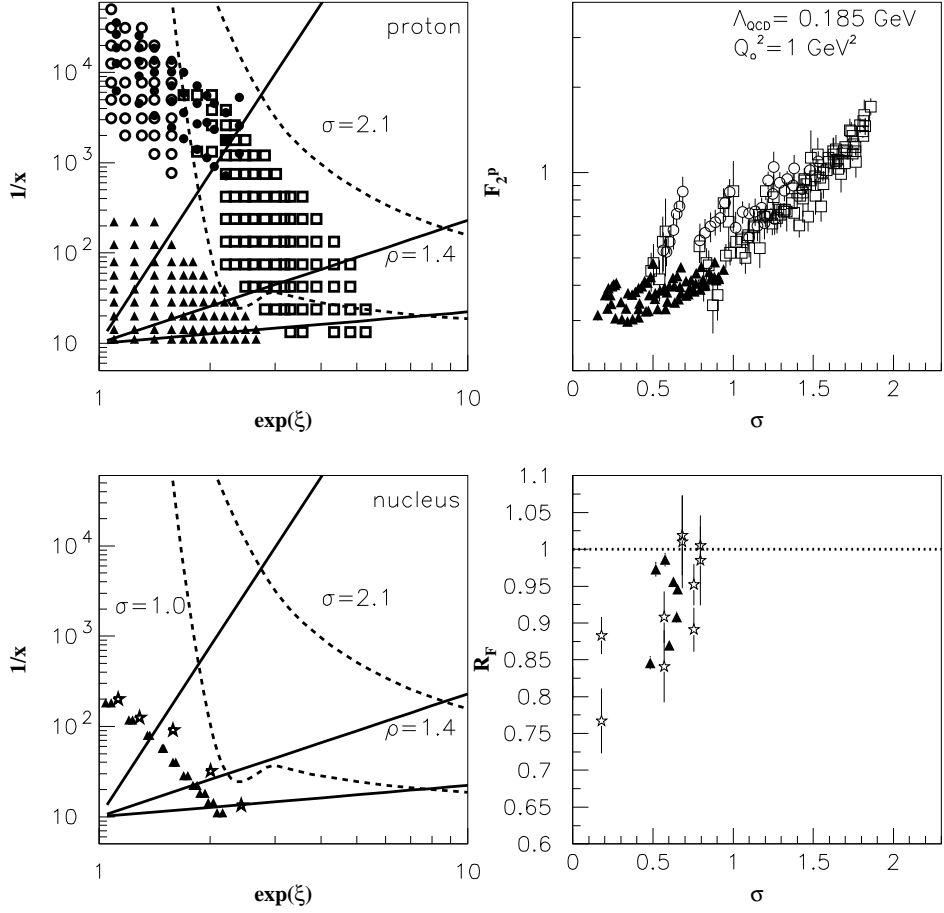,width=14.truecm}}
\vskip0.6truecm
\caption{The asymptotic region in the proton (top) and in a nucleus (bottom).
The figures on the left show the kinematical range in $Y=1/x$ and 
$\xi$ covered by current experiments: NMC \protect\cite{NMC} \protect\cite{NMC1} (triangles),
H1 \protect\cite{H1}a (squares) and \protect\cite{H1}b,c (full dots), 
ZEUS \protect\cite{ZEUS} (open dots), E665 \protect\cite{E665}
(stars). The asymptotic region is delimited by the values: 
$0.7 < \rho < 3$ (full lines) and 
$1 < \sigma < 2.1$ (dashed lines). 
The figures on the right show the onset of DAS in the proton (top) 
and the shadowing ratio, $R_F$, vs. $\sigma$ in a nucleus (bottom). 
While DAS is achieved in the asymptotic region covered mostly by the
H1 \protect\cite{H1} data, a regular pattern is yet to be seen in the nuclear data which
are clearly lying outside the asymptotic region (left bottom).}
\label{kin}
\end{figure}

\begin{figure}[htb]
%\vspace{9pt}${40}Ca$
%\framebox[55mm]{\rule[-21mm]{0mm}{43mm}}
%  \leavevmode
%\epsfxsize=3.5in
%\epsfysize=2.0in
%\epsffile{ndep3.eps}  
\vbox{
\hskip.6truecm\epsfig{figure=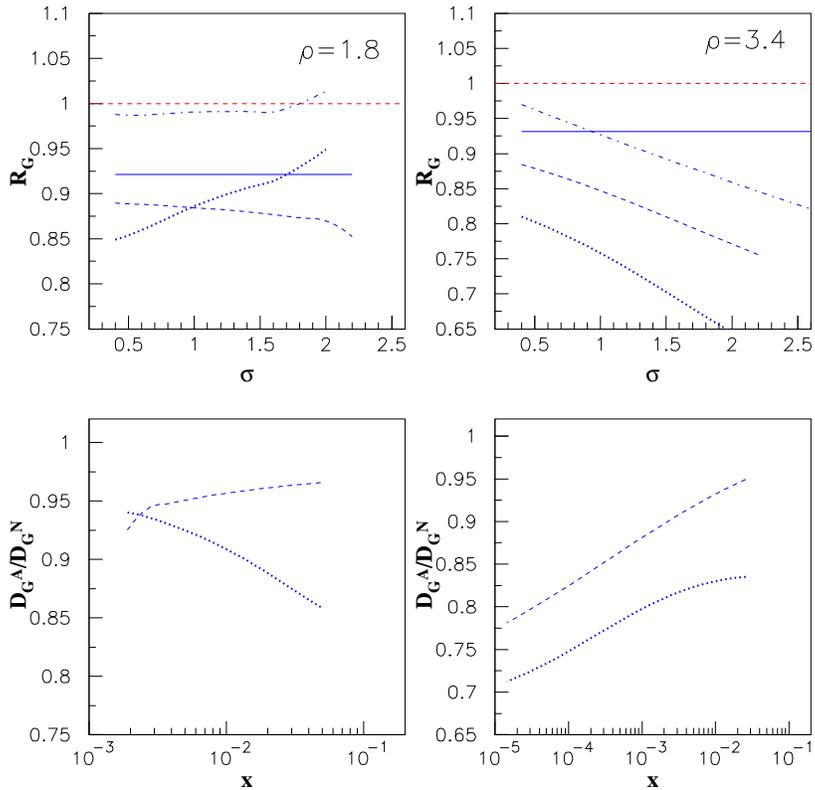,width=12.truecm}}
\vskip0.6truecm
\caption{Top: Ratio $R_G=G_A/G_N$, vs. 
DAS variable $\sigma$ at two different values of $\rho$: $\rho=1.8$ (left) 
and $\rho=3.4$ (right).     
The theoretical curves show both $\sigma$-scaling (full lines) and the $\sigma$-scaling
violations predicted in this paper: dashes, Eq.(15a), dots, Eq.(15b), 
and dot-dashes, Eq.(15c). 
Bottom: Ratio of nuclear to nucleon damping factors (see Eq.(\protect\ref{GSC})), as a function of $x$ 
for the same values of $\rho$ and $\sigma$ used in the shadowing calculations above.
The $Q^2$ ranges are: $2.3 < Q^2 < 2 \times 10^3 \, {\rm GeV^2}$ (left),  
$1.5 < Q^2 < 40. < \, {\rm GeV^2}$ (right). 
We show results using the AJM 
of Ref.\protect\cite{FS} as our non-perturbative input; 
analogous scaling relations are found by using other initial 
shadowing models.}
%\vskip-.8truecm
\label{scaling}
\end{figure}

\end{document}